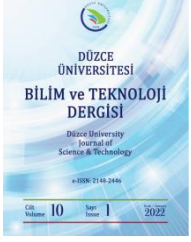

# Düzce University
# Journal of Science & Technology

*Research Article*

# Tooth Instance Segmentation on Panoramic Dental Radiographs Using U-Nets and Morphological Processing

Selahattin Serdar HELLİ [a],  Andaç HAMAMCI [a,*]

[a] *Department of Biomedical Engineering, Faculty of Engineering, Yeditepe University, İstanbul, TURKEY*
*\* Corresponding author's e-mail address: andac.hamamci@yeditepe.edu.tr*
DOI: 10.29130/dubited.950568

## ABSTRACT

Automatic teeth segmentation in panoramic x-ray images is an important research subject of the image analysis in dentistry. In this study, we propose a post-processing stage to obtain a segmentation map in which the objects in the image are separated, and apply this technique to tooth instance segmentation with U-Net network. The post-processing consists of grayscale morphological and filtering operations, which are applied to the sigmoid output of the network before binarization. A dice overlap score of 95.4±0.3% is obtained in overall teeth segmentation. The proposed post-processing stages reduce the mean error of tooth count to 6.15%, whereas the error without post-processing is 26.81%. The performances of both segmentation and tooth counting are the highest in the literature, to our knowledge. Moreover, this is achieved by using a relatively small training dataset, which consists of 105 images. Although the aim in this study is to segment tooth instances, the presented method is applicable to similar problems in other domains, such as separating the cell instances.

# U-Net ve Morfolojik İşlemler Kullanılarak Panoramik Radyografiler Üzerinde Diş Örneği Bölütleme

## ÖZ

Panoramik röntgen görüntülerinde otomatik diş bölütleme, diş hekimliği görüntü analizinin önemli bir araştırma konusudur. Bu çalışmada, görüntüdeki nesnelerin ayrıldığı bir bölütleme haritası elde etmek için bir son işleme aşaması öneriyoruz ve bu tekniği U-Net ağı ile diş örneği bölütlemeye uyguluyoruz. Son işleme, ikililleştirmeden önce ağın sigmoid çıkışına uygulanan gri tonlamalı morfolojik ve filtreleme işlemlerinden oluşmaktadır. Tüm diş bölütlemede %95,4±0,3'lük bir Dice örtüşme puanı elde edilmiştir. Önerilen son işleme aşamaları, diş sayısının tespitinde ortalama hatayı %26,81'den %6,15'e düşürmüştür. Bildiğimiz kadarıyla hem bölütleme, hem de diş sayma performansları literatürdeki en yüksek performanslardır. Ayrıca bu sonuç, 105 görüntüden oluşan nispeten küçük bir eğitim veri seti kullanılarak elde edilmiştir. Bu çalışmadaki amaç diş örneklerini bölütlemek olsa da, sunulan yöntem hücre örneklerini ayırmak gibi diğer alanlardaki benzer problemlere uygulanabilir.

*Anahtar Kelimeler:* Diş, Panoramik, Bölütleme, Örnek bölütleme, Sayma, U-Net

# I. INTRODUCTION

Nowadays, oral health is an important indicator of overall health, and quality of life. Oral health is multifaceted and includes the ability to speak, smile, smell, taste, touch, chew, swallow, and convey a range of emotions through facial expressions with confidence and without pain, discomfort, and disease of the craniofacial complex [1]. Major oral diseases include dental caries (tooth decay), periodontal (gum) disease, and oral cancers. Although those diseases are largely preventable, oral diseases are highly prevalent conditions, affecting more than 3.5 billion people around the world. Dental caries is the most common disease globally with increasing prevalence [2]. For the diagnosis of oral diseases, radiographs are valuable tools, supporting the clinical examination. Periapical radiographs and panoramic radiographs are routinely used and provide information necessary for routine dental practice. However, projection of the entire or some part of the mouth onto two dimensional image plane has certain limitations and in diagnosis or treatment planning of special cases, advanced three-dimensional imaging modalities, revealing additional information is desirable. Three-dimensional imaging techniques that are applied in dentistry include cone beam computed tomography (CBCT), computed tomography (CT), magnetic resonance imaging (MRI) and ultrasonography [3].

Despite the aforementioned disadvantage, with the wide availability, lower dose of ionizing radiation (compared to CBCT), and patient comfort, panoramic dental x-ray is a commonly performed medical examination by dentists and oral surgeons in everyday practice and is an important diagnostic tool. Panoramic radiographs provide a broad overview of the orofacial region including jaws, teeth, sinuses and temporomandibular joint (TMJ). They are especially useful in showing dental development stages or anomalies, or as an initial examination for generalized disease or multiple problems [4].

Segmentation of teeth is usually a necessary step in the analysis of dental images such as for lesion detection [5], age or gender determination [6] and human identification [7]. Automatic teeth segmentation in panoramic x-ray images is an important research subject of the image analysis in oral medicine. To isolate teeth on panoramic radiographs is challenging, since radiographs show other parts of the patient's body (e.g., chin, spine and jaws) [8]. An in-depth review of segmentation methods applied to panoramic radiographs, which do not use machine learning techniques, is provided in [9]. Deep learning algorithms, in particular convolutional neural networks (CNNs), have rapidly become a methodology of choice for analyzing medical images with the advent of computation hardware/algorithms, increase in the amount of data and superior success [10]. In the literature, various studies utilize convolutional neural networks for segmentation and identification of teeth in dental images.

In [11], Zhao et.al. proposed a two-stage network, which they called TSASNet, to segment teeth on dental panoramic X-ray images. The first stage was an attention model to roughly localize the tooth region, whereas a fully convolutional network was employed to obtain a fine segmentation at the second stage. They reported 92.72% average dice overlap on the dataset of [9] that consists of 1500 dental panoramic images.

Another group of study aims to detect and identify each tooth on dental images, instead of segmenting. Faster regional convolutional networks (faster R-CNN) [12] are often employed for this purpose. In [13], using faster R-CNN, followed by post-processing techniques according to certain prior domain knowledge was proposed to detect and identify teeth in dental periapical films. In [14], a faster R-CNN and heuristic methods were applied to detect and number the teeth and implants on dental panoramic radiograms. The accuracy of tooth numbering was reported as 84.5%. In [15], a two stage system was proposed, in which faster R-CNN is used to detect the teeth followed by a VGG-16 network [16] to identify and number. In [17], to identify the teeth on dental periapical radiographs, due to insufficient amount of data, a label tree was used with neural networks.

A related problem, tooth instance segmentation problem, which refers to segmenting each tooth in the image separately, attracts interest of researchers in dental image analysis area. This is usually performed by consecutive application of detection and segmentation, and often by using Mask R-CNN structure [18]. Utilizing Mask R-CNN on dental panoramic radiograms, Jader et.al. reported 88% F1-score [8], whereas Lee et.al. reported 87.5% F1-score [19]. Both studies have initialized the models with pre-trained weights. In [20], a two stage pipeline was proposed on panoramic radiograms, where the first deeplab (v3) network [21] detects and classifies the teeth, whereas the second fully convolutional neural network performs the segmentation. Similar techniques were also applied for instance segmentation on CBCT volumes [22,23] and on intra-oral optical scans [24].

In one of the first applications of popular U-Net network [25], Ronneberger et.al. performed semantic segmentation of tooth in dental x-ray images into the classes caries, enamel, dentin, pulp, crown, restoration and root canal treatment [26]. With the U-Net network, Koch et.al. reached a Dice score of 93.4% for the teeth segmentation on panoramic radiographs [27]. This result obtained with simpler U-Net architecture was better at that time, when compared to the performance of more sophisticated models, tested on the same dataset. However, the output of the network is an overall segmentation map, instead of split teeth segmentations.

In this study, we propose to apply a post-processing stage to obtain a segmentation map in which the objects in the image are separated and use the technique for tooth instance segmentation with U-Net network. The post-processing operations consist of grayscale morphological and filtering operations, which are applied to the sigmoid output of the network, before binarization. The proposed post-processing stage is inspired of the method proposed by Lu and Ke to separate touching round shaped objects in photographs [28]. Their method to remove small light details was mainly based on opening operations of gray-scale morphology and a very high success in separation while preserving the shape of the objects was reported. For a similar problem in another domain, segmentation of each cell instance separately, Falk et.al. proposed to modify loss computation of U-Net. An artificial 1-pixel-wide background ridge between touching instances were inserted in the segmentation mask and their weight in the loss computation were increased, such that the thinnest ridges have the highest weights [29].

The aim of this study is to develop a method based on U-Net network and morphological processing to perform tooth instance segmentation on panoramic dental radiographs in order to provide diagnostic information for the management of dental disorders, diseases, and conditions.

## II. METHODS

### A. DATASET

The dataset, which was used in [30], consists of anonymized and de-identified panoramic dental x-ray images of 116 patients, taken at Noor Medical Imaging Center, Qom, Iran. The manual segmentations of mandibles are also available in the dataset but those segmented images are irrelevant to the subject of our study and only original images are used. The images have been taken by the Soredex CranexD digital panoramic x-ray unit. The widths of all images vary between 2600-3138 pixels, their heights are between 1050-1380 pixels. Complete edentulous cases are excluded in this study.

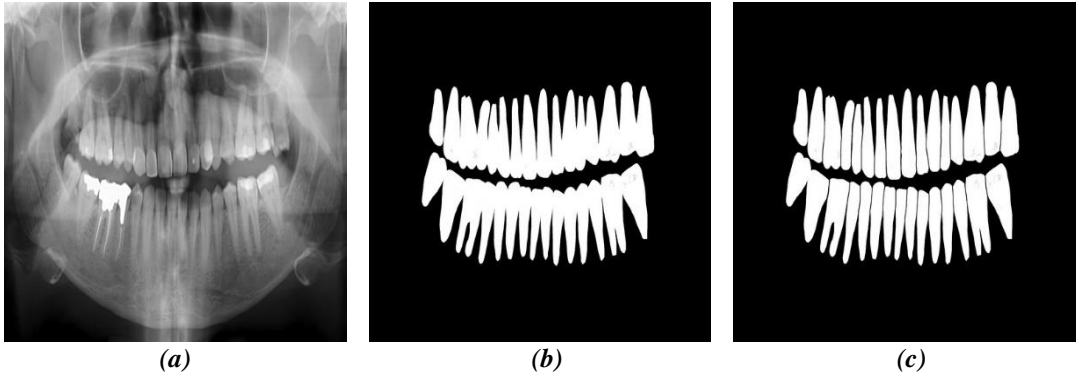

***Figure 1.*** *Examples of masks which have been used in this study.* ***(a)*** *The original radiograph.* ***(b)*** *Full mask obtained by manual labeling.* ***(c)*** *Split mask, in which each tooth is separated from the others by a narrow gap.*

During data preprocessing, all panoramic dental x-ray images are resized to 512x512 pixels, and normalized in the range of 0 to 1. An example input image is given in Figure 1 (a). For each panoramic dental x-ray image in the dataset, two different teeth masks are obtained by manual labeling. In the first mask, all teeth are labeled as given in Figure 1 (b). For the second one, although the teeth are arranged in contact, each tooth is separated from the others by a narrow gap in between as given in Figure 1 (c).

## B. THE NETWORK ARCHITECTURE

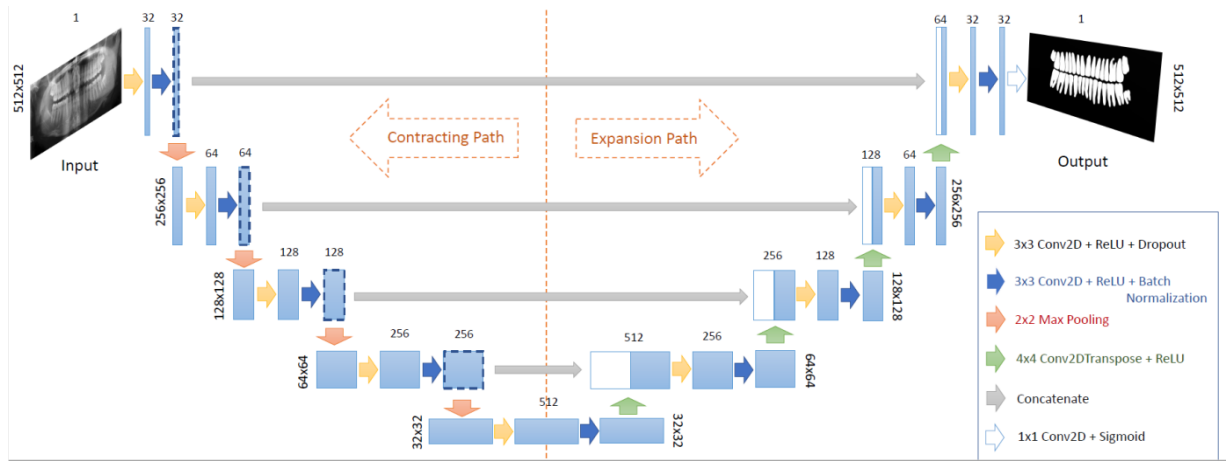

***Figure 2.*** *The U-Net architecture used in the study. Each blue box corresponds to a multi-channel feature map. The number of channels is denoted on top of the box. The shape is provided on the edge of the box. White boxes represent copied feature maps. The arrows denote the different operations indicated by colors.*

The U-net architecture [25], which is a fully convolutional neural network and popular in biomedical image segmentation, is used in this study. The network architecture is illustrated in Figure 2. The U-net is a symmetric architecture consists of an encoder network, which maps the image into lower-dimensional latent representation, followed by a decoder network, which reconstructs the output by up-sampling the latent vector back to the input size.

At each level of the contracting path, two convolutional layers with 3x3 kernels and rectified linear unit (ReLU) activation functions, followed by a batch normalization, are applied. The feature maps are downsampled by a factor of 2, whereas the number of features are doubled, by 2x2 max pooling operations at each step. In the expanding path, the upsampling is performed by 4x4 transposed convolution. The rate of dropout operations applied at each level are 0.15, 0.2, 0.3, 0.4, 0.5, 0.4, 0.3, 0.2, 0.1, in the order from the input to the output level. By the skip connections, the features are transferred from each level of the contracting path to the same level of the expanding path. Two convolutional layers with 3x3 kernels and ReLU activation functions, followed by a batch normalization, are applied after each upscaling operation. In the final step, the output of the network is produced by applying a 1x1 convolution and a sigmoid activation function.

## C. TRAINING DETAILS

The network is implemented in Python using Keras library. The source code, as well as the manual segmentations are available online in Github repository (https://github.com/ImagingYeditepe/Segmentation-of-Teeth-in-Panoramic-X-ray-Image). The loss function is binary cross entropy. The weights have been initialized randomly by sampling from a truncated normal distribution centered at 0, as explained in [31]. The weights are optimized through the Adaptive Moment Estimation (ADAM) optimizer [32] at 250 epochs with a batch size of 4. The learning rate is 0.001. In this study, different combinations of horizontal flipping, vertical flipping, and adding random salt and pepper noise are applied for data augmentation to the training dataset as explained in Section E. For the salt and pepper noise, the proportion of image pixels to replace with noise on range [0,1] is 5%.

## D. POST-PROCESSING

Instead of an immediate binarization of the sigmoid output of the network, separation of tooth instances in the final map is assured by morphological operations. For this purpose, each output map produced by the network is processed by a series of basic image processing operations of the "Open Computer Vision (OpenCV)" library. The aim of these operations is not only to increase the separation of the teeth from each other, but also to reduce the noise in the final segmentation maps. A diagram of the post-processing steps is given in Figure 3.

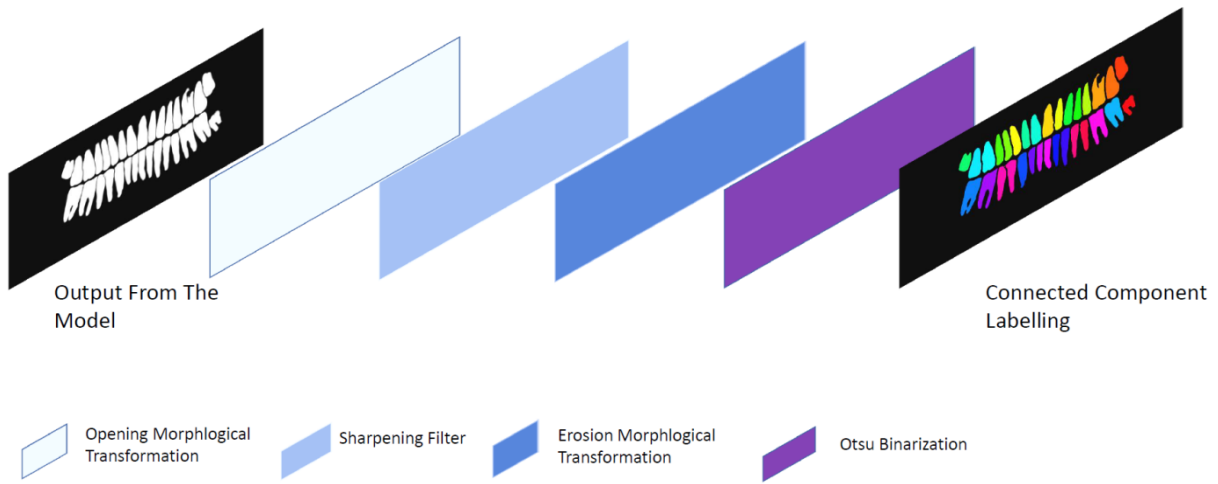

*Figure 3.* Steps of the post-processing that are applied to the network output.

First, the output of the network has been resized to the size of the original input, using a Lanczos filter (a truncated sinc) on all pixels that may contribute to the output value. Secondly, morphological grayscale opening operation on resized output map is applied to remove small light details in the image and to separate teeth from each other [28]. The square shaped structural element used in all morphological operations is a 5x5 matrix which consists of ones. Then, the image sharpening filter with the kernel given in Eqn. 1 is applied to further enhance the details.

$$k = \begin{bmatrix} -1 & -1 & -1 \\ -1 & 9 & -1 \\ -1 & -1 & -1 \end{bmatrix} \tag{1}$$

Before the segmentation, to detach the connected teeth, grayscale erosion morphological operation is applied twice to the output map of the sharpening filter. Square shaped structural element, the same as the opening, is used. Following the erosion operation, the masks are generated by a segmentation operation, in which the optimum threshold is determined by the Otsu's method [33]. Finally, subsets of connected components are uniquely labeled on the masks using the connected-component procedure, with a cluster size threshold of 2000 pixels.

## E. EXPERIMENTS AND PERFORMANCE EVALUATION

*Table 1. Experiments*

| # | AUGMENTATION | | | TRAINING OUTPUT | POST PROC |
| | *S&P Noise* | *Horizontal Flip* | *Vertical Flip* | | |
|---|---|---|---|---|---|
| E1 | - | - | - | Full Mask | - |
| E2 | - | - | - | Split Mask | - |
| E3 | YES | YES | YES | Split Mask | - |
| E4 | YES | YES | - | Split Mask | - |
| E5 | - | - | - | Split Mask | YES |

Five experiments are performed to determine the performance of the proposed method under various parameters including different output labels, data augmentation or post processing scenario. In these experiments, a 10-fold cross validation technique is applied. The comparison of the performed procedures is given in Table 1. Whenever the post-processing steps are not applied, segmentation masks are obtained by directly binarizing the sigmoid output of the network using a threshold level of 0.2.

- The First Experiment (E1): In the E1, the baseline performance of the U-Net is evaluated. The original images and full segmentation masks of teeth are used to train the network, without any data augmentation nor post processing.
- The Second Experiment (E2): E2 is performed to evaluate the effect of introducing gaps between the teeth. The training dataset consists of the original images and segmentation masks of teeth in which each tooth has been separated from each other. No data augmentation nor post-processing is applied.
- The Third Experiment (E3): To understand the value of data augmentation in training the network, data augmentation step, which consists of vertical flipping, horizontal flipping, and adding random noise techniques is applied to the input images of E2.
- The Fourth Experiment (E4): Panoramic radiographs has a horizontal symmetry. Hence applying an augmentation by flipping the images on horizontal axis is natural. However the same can be said for vertical flipping. For this reason, the training of E3 is repeated without applying vertical flipping in the augmentation.
- The Fifth Experiment (E5): Finally, in the fifth experiment, the effect of post-processing operations to separate teeth is compared to the result of E2. The training is the same as E2, but proposed post-processing operations are applied in the testing stage.

To evaluate the segmentation results; specificity, sensitivity, positive predictive value (PPV), negative predictive value (NPV), dice similarity coefficient (Dice) and jaccard index (Jaccard) are calculated as given in Eqn. 2. In these equations; TP, TN, FN, and FP stand for true positive, true negative, false negative, and false positive, respectively. These metrics are used in a pixel-wise fashion.

$$\text{Sensitivity} = \frac{TP}{TP + FN} \qquad \text{Specificity} = \frac{TN}{TN + FP} \qquad \text{PPV} = \frac{TP}{TP + FP}$$

$$\text{NPV} = \frac{TN}{TN + FN} \qquad \text{Jaccard} = \frac{TP}{TP + FP + FN} \qquad \text{Dice} = \frac{2TP}{2TP + FP + FN} \qquad (2)$$

To evaluate the connected-component labeling results, mean absolute percentage error (MAPE) metric is used as given in Eqn. 3. In this equation; N is the number of images, the actual value is the number of teeth in the image and the predicted value is the estimated number of teeth by the proposed system.

$$Err = \frac{1}{N} \sum_{1}^{N} \frac{|Actual\ Value - Predicted\ Value|}{|Actual\ Value|} \times 100 \qquad (3)$$

The average and standard deviation of each metric for the cross-validation folds are calculated and the statistical significance of the difference in fold averages between various approaches are assessed using the Wilcoxon signed-rank test with a significance level of 0.01 [34].

# III. RESULTS

***Table 2.*** *Results of the experiments.*

|     | Sensitivity | Specificity | PPV | NPV | Jaccard | Dice |
|-----|-------------|-------------|-----|-----|---------|------|
| E1 | 96.6±0.9 | 98.4±0.4 | 92.6±1.9 | 99.3±0.2 | 89.7±1.2 | 94.5±0.7 |
| E2 | 96.3±0.6 | 98.7±0.3 | 93.9±1.0 | 99.2±0.1 | 90.6±0.8 | 95.1±0.5 |
| E3 | 96.1±0.6 | 98.8±0.3 | 94.4±0.8 | 99.2±0.1 | 90.9±0.6 | 95.2±0.3 |
| E4 | 95.9±0.6 | 98.9±0.2 | 94.8±0.8 | 99.1±0.2 | 91.1±0.5 | 95.4±0.3 |
| E5 | 99.1±0.3 | 98.6±0.2 | 91.9±1.2 | 99.9±0.0 | 91.1±1.2 | 95.3±0.6 |

The values are given as percentages in mean ± std.dev. format

The quantitative results of the five experiments described in Section II-E are given in Table 2.

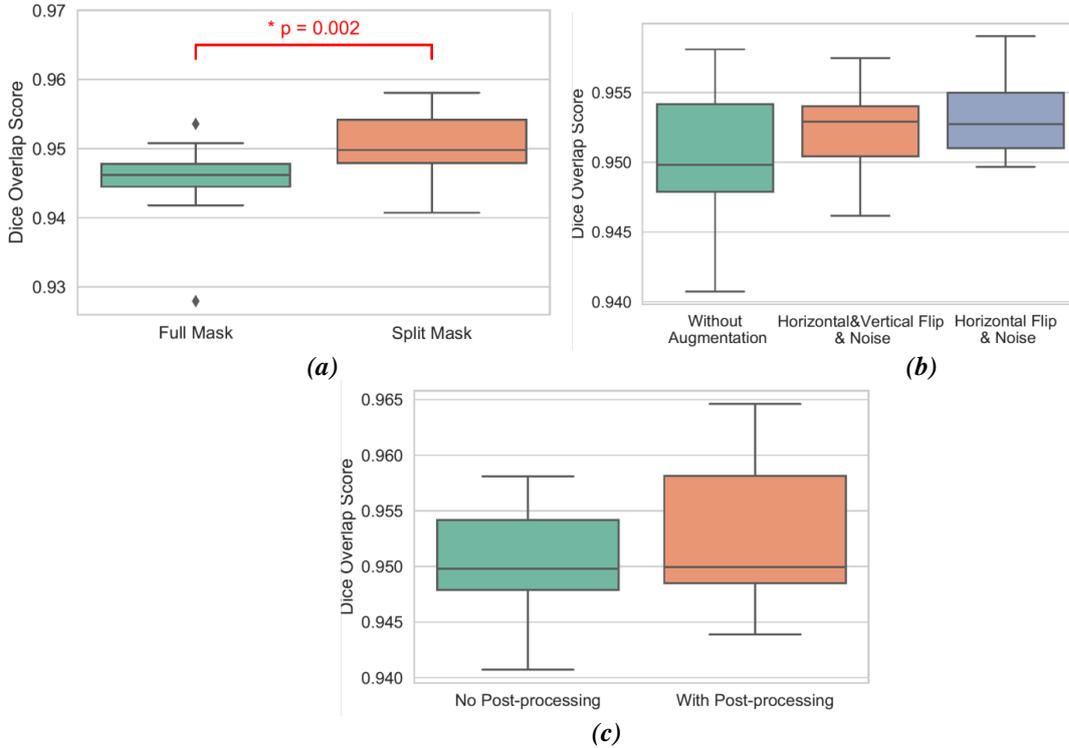

***Figure 4.*** *(a) Comparison of the segmentation performance of the network on the full mask (E1) and on the split mask (E2), without any augmentation nor post processing. (b) Comparison of the segmentation performance of the network with three different augmentation strategies (E2, E3 and E4). (c) Comparison of the segmentation performance after post-processing with the proposed stages (E5) and without post-processing (E2).*

45

To understand the effect of artificially splitting each tooth on the output mask, the segmentation is performed on the full mask in E1 and on the split mask in E2, without any augmentation nor post processing operations. The distribution of the dice scores obtained with the full mask (E1) and with the split mask (E2) are given as a box plot in Figure 4 (a). The dice overlap score is significantly (p = 0.002) higher for the split mask compared to the full mask.

Next, to evaluate the value of data augmentation applied on the training dataset, the segmentation accuracy with two different data augmentation strategies are compared to the baseline network. The distribution of dice scores for three experiments are presented with the box plots in Figure 4 (b). Although a slight shift towards higher dice scores is observed by the training data augmentation using the horizontal flipping and addition of noise, the difference is not significant (p = 0.049). However, addition of vertical flipping does not improve the performance (p=0.275).

Finally, to demonstrate the effect of post-processing stages on the segmentation performance, the dice score after post-processing is compared with the simple thresholding approach and plotted in Figure 4 (c). The slight shift towards the higher dice score is not significant (p = 0.037).

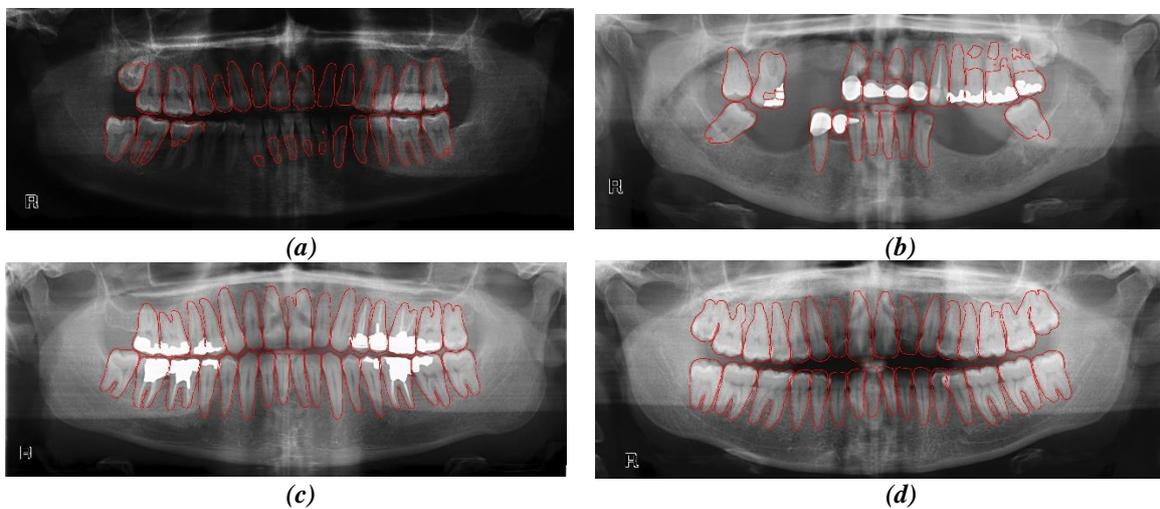

*(a)*                     *(b)*

*(c)*                     *(d)*

***Figure 5.*** *Examples of segmentation results. **(a)** This output of the model has the worst segmentation map. However, as can be seen, the original image has poor diagnostic information. **(b)** This example of the model's output has a poor segmentation map. **(c,d)** Examples of the model's output having an effective segmentation map.*

Examples of segmentation results are presented qualitatively in Figure 5.

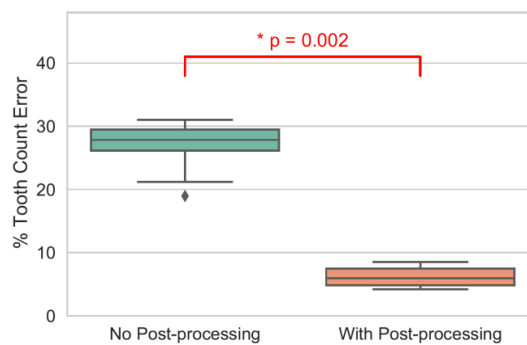

***Figure 6.*** *Comparison of the error in tooth count after post-processing with the proposed stages and without post-processing (E2 and E5).*

The mean error of tooth count when the proposed post-processing stages applied is 6.15%, whereas the error without post-processing is 26.81%. This error is significantly different in folds between two approaches (p = 0.002). The error in estimating the number of teeth in panoramic images after the post-processing stages and without post-processing is given in the boxplot in Figure 6.

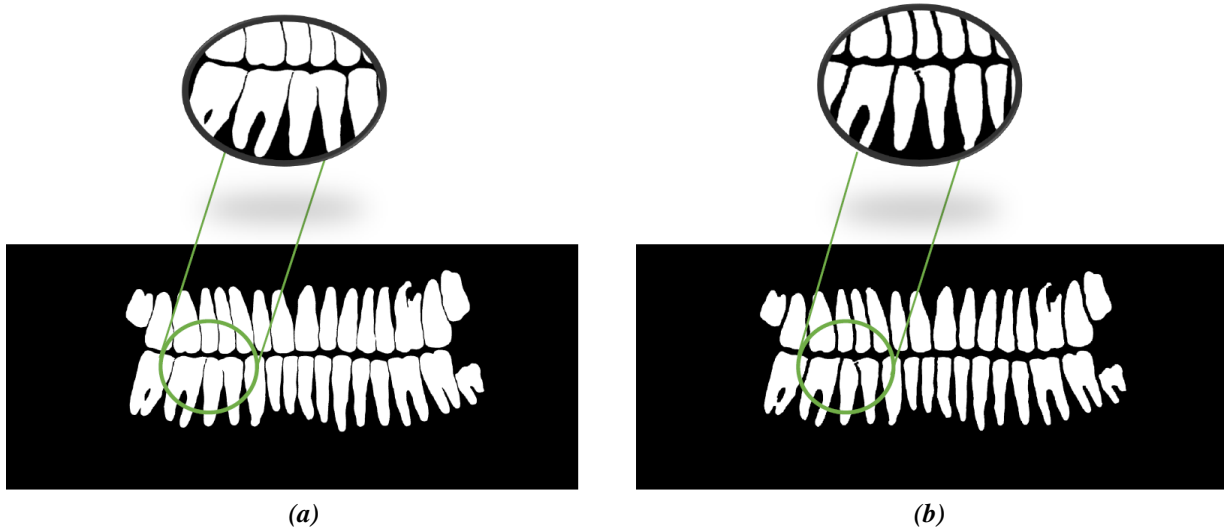

<div align="center">(<b>a</b>)                                  (<b>b</b>)</div>

***Figure 7.*** *An example output of the network (**a**) without post-processing operations and (**b**) with the proposed post-processing operations.*

The effect of post-processing operations on the segmentation result is presented on a sample output in Figure 7.

# IV. DISCUSSION

In this study, a method to segment the tooth instances on panoramic images is proposed. The separation of teeth is provided by application of post-processing operations to the output of the sigmoid classification layer of the neural network.

To compare the proposed method with the relevant literature, we first used the full segmentation maps, without separation of tooth instances. In the literature, one of the highest successes to segment the teeth in the panoramic radiographs was reported by using the U-Net neural networks [27]. The 94.5% dice score obtained on fully annotated mask using U-Net in the present study is similar to 92.8% reported in [27]. Slightly better result can be due to the difference in image sets or annotations.

Although the arrangement of the teeth are contiguous on panoramic radiographs, we artificially introduced gaps between the teeth on the ground truth label maps to separate the instances. Still, the segmentation performance on the split mask is even higher than that on the full mask.

The number of radiographs used in training the network is lower, compared to the related studies in the literature. In [8], one of the smallest datasets that consists of 193 annotated images was used to train the Mask R-CNN. Due to the number of free parameters of the network, the data were not sufficient to train the network from scratch, and pre-trained weights were taken from other resources. In this study, we trained the network from scratch by initializing the weights randomly. The main reason allowing this is that the separation of instances is performed by morphological operations, instead of a trainable network. To further compensate the dataset limitation, data augmentation strategies were applied. Adding salt and pepper noise and horizontal flipping resulted a slight increase in the performance. However, a flipping around the vertical axis had a negative effect on the performance, probably due to the asymmetry of the panoramic radiographs in superior-inferior direction.

The performance of counting the number of teeth is significantly improved by the application of the post-processing stages. In [35], tooth instance segmentation is performed on dental CT images by grayscale morphological operations, followed by a watershed segmentation. They reported 10% error in estimating the number of teeth compared to the actual number of teeth. In [14], 84.5% accuracy is reported in numbering the teeth on panoramic radiograms with a faster R-CNN structure. In our study, by the proposed technique, the mean error of tooth count is 6.15% which is the lowest in the literature, to our knowledge. The post processing operations, that are applied to the sigmoid output of the network not only reduced the error in counting the teeth, but also significantly increased the dice performance of the final segmentation map.

# V. CONCLUSION

In this study, a method to segment and count the tooth instances on panoramic radiographs is proposed. The evaluations demonstrate the potential of the method to help clinical practice by serving as a prior step for further processing and analysis of dental images. The performances of both segmentation and tooth counting are the highest in the literature, to our knowledge. Moreover, this is achieved by using a relatively small training dataset, which consists of 105 images. The technique proposed in this study is based on image processing stages that are applied to the sigmoid output of the neural network before binarization. Although, the aim in this study is to segment tooth instances, the presented method is applicable to similar problems on other domains, such that separating the cell instances.

ACKNOWLEDGEMENT:    The authors thank Ayşe Merve Kablan and Ahmet E. Yetkin for helpful comments.

# VI. KAYNAKLAR

[1]     M. Glick, D. M. Williams, D. V. Kleinman, M. Vujicic, R. G. Watt, and R. J. Weyant, "A new definition for oral health developed by the fdi world dental federation opens the door to a universal definition of oral health," *British dental journal*, vol. 221, no. 12, pp. 792–793, 2016.

[2]     M. A. Peres, L. M. Macpherson, R. J. Weyant, B. Daly, R. Venturelli, M. R. Mathur, S. Listl, R. K. Celeste, C. C. Guarnizo-Herreño, C. Kearns et al., "Oral diseases: a global public health challenge," *The Lancet*, vol. 394, no. 10194, pp. 249–260, 2019.

[3]     N. Shah, N. Bansal, and A. Logani, "Recent advances in imaging technologies in dentistry," *World journal of radiology*, vol. 6, no. 10, p. 794, 2014.

[4]     B. Vandenberghe, R. Jacobs, and H. Bosmans, "Modern dental imaging: a review of the current technology and clinical applications in dental practice," *European radiology*, vol. 20, no. 11, pp. 2637–2655, 2010.

[5]     R. G. Birdal, E. Gumus, A. Sertbas, and I. S. Birdal, "Automated lesion detection in panoramic dental radiographs," *Oral Radiology*, vol. 32, no. 2, pp. 111–118, 2016.

[6]     E. Avuclu and F. Basciftci, "Novel approaches to determine age and gender from dental x-ray images by using multiplayer perceptron neural networks and image processing techniques," *Chaos, Solitons & Fractals*, vol. 120, pp. 127 – 138, 2019.

[7]     M. H. Bozkurt and S. Karagol, "Jaw and teeth segmentation on the panoramic x-ray images for dental human identification," *Journal of Digital Imaging*, pp. 1–18, 2020.

[8]     G. Jader, J. Fontineli, M. Ruiz, K. Abdalla, M. Pithon, and L. Oliveira, "Deep instance segmentation of teeth in panoramic x-ray images," in *2018 31st SIBGRAPI Conference on Graphics, Patterns and Images (SIBGRAPI)*. IEEE, 2018, pp. 400–407.

[9]     G. Silva, L. Oliveira, and M. Pithon, "Automatic segmenting teeth in x-ray images: Trends, a novel data set, benchmarking and future perspectives," *Expert Systems with Applications*, vol. 107, pp. 15–31, 2018.

[10]     G. Litjens, T. Kooi, B. E. Bejnordi, A. A. A. Setio, F. Ciompi, M. Ghafoorian, J. A. Van Der Laak, B. Van Ginneken, and C. I. Sánchez, "A survey on deep learning in medical image analysis," *Medical image analysis*, vol. 42, pp. 60–88, 2017.

[11]     Y. Zhao, P. Li, C. Gao, Y. Liu, Q. Chen, F. Yang, and D. Meng, "Tsasnet: Tooth segmentation on dental panoramic x-ray images by two-stage attention segmentation network," *Knowledge-Based Systems*, vol. 206, p. 106338, 2020.

[12]     S. Ren, K. He, R. Girshick, and J. Sun, "Faster r-cnn: Towards real-time object detection with region proposal networks," in *Advances in neural information processing systems*, 2015, pp. 91–99.

[13]     H. Chen, K. Zhang, P. Lyu, H. Li, L. Zhang, J. Wu, and C.-H. Lee, "A deep learning approach to automatic teeth detection and numbering based on object detection in dental periapical films," *Scientific reports*, vol. 9, no. 1, pp.1–11, 2019.

[14]     C. Kim, D. Kim, H. Jeong, S.-J. Yoon, and S. Youm, "Automatic tooth detection and numbering using a combination of a cnn and heuristic algorithm," *Applied Sciences*, vol. 10, no. 16, p. 5624, 2020.

[15]     D. V. Tuzoff, L. N. Tuzova, M. M. Bornstein, A. S. Krasnov, M. A. Kharchenko, S. I. Nikolenko, M. M. Sveshnikov, and G. B. Bednenko, "Tooth detection and numbering in panoramic radiographs using convolutional neural networks," *Dentomaxillofacial Radiology*, vol. 48, no. 4, p. 20180051, 2019.

[16]     K. Simonyan and A. Zisserman, "Very deep convolutional networks for large-scale image recognition," *arXiv preprint arXiv:1409.1556*, 2014.

[17]     K. Zhang, J. Wu, H. Chen, and P. Lyu, "An effective teeth recognition method using label tree with cascade network structure," *Computerized Medical Imaging and Graphics*, vol. 68, pp. 61–70, 2018.

[18]     K. He, G. Gkioxari, P. Dollár, and R. Girshick, "Mask r-cnn," in *Proceedings of the IEEE international conference on computer vision*, 2017, pp. 2961–2969.

[19]     J.-H. Lee, S.-S. Han, Y. H. Kim, C. Lee, and I. Kim, "Application of a fully deep convolutional neural network to the automation of tooth segmentation on panoramic radiographs," *Oral surgery, oral medicine, oral pathology and oral radiology*, vol. 129, no. 6, pp. 635–642, 2020.

[20]     A. F. Leite, A. Van Gerven, H. Willems, T. Beznik, P. Lahoud, H. Gaêta-Araujo, M. Vranckx, and R. Jacobs, "Artificial intelligence-driven novel tool for tooth detection and segmentation on panoramic radiographs," *Clinical Oral Investigations*, pp. 1–11, 2020.

[21]     L.-C. Chen, G. Papandreou, I. Kokkinos, K. Murphy, and A. L. Yuille, "Deeplab: Semantic image segmentation with deep convolutional nets, atrous convolution, and fully connected crfs," *IEEE transactions on pattern analysis and machine intelligence*, vol. 40, no. 4, pp. 834–848, 2017.

[22]    M. Chung, M. Lee, J. Hong, S. Park, J. Lee, J. Lee, I.-H. Yang, J. Lee, and Y.-G. Shin, "Pose-aware instance segmentation framework from cone beam ct images for tooth segmentation," *Computers in Biology and Medicine*, pp. 103720, 2020.

[23]    Z. Cui, C. Li, and W. Wang, "Toothnet: Automatic tooth instance segmentation and identification from cone beam ct images," in *Proceedings of the IEEE Conference on Computer Vision and Pattern Recognition*, 2019, pp.6368–6377.

[24]    F. G. Zanjani, D. A. Moin, F. Claessen, T. Cherici, S. Parinussa, A. Pourtaherian, S. Zinger et al., "Mask-mcnet: Instance segmentation in 3d point cloud of intra-oral scans," in *International Conference on Medical Image Computing and Computer-Assisted Intervention*. Springer, 2019, pp. 128–136.

[25]    O. Ronneberger, P. Fischer, and T. Brox, "U-net: Convolutional networks for biomedical image segmentation," in *International Conference on Medical image computing and computer-assisted intervention*, Springer, 2015, pp.234–241.

[26]    O. Ronneberger, P. Fischer, and T. . Brox, "Dental x-ray image segmentation using a u-shaped deep convolutional network," in *International Symposium on Biomedical Imaging (ISBI)*, 2015.

[27]    T. L. Koch, M. Perslev, C. Igel, and S. S. Brandt, "Accurate segmentation of dental panoramic radiographs with u-nets," in *2019 IEEE 16th International Symposium on Biomedical Imaging (ISBI 2019)*, 2019, pp. 15–19.

[28]    N. Lu and X. Ke, "A segmentation method based on gray-scale morphological filter and watershed algorithm for touching objects image," in *Fourth International Conference on Fuzzy Systems and Knowledge Discovery (FSKD 2007)*, vol. 3, 2007, pp. 474–478.

[29]    T. Falk, D. Mai, R. Bensch, Ö. Çiçek, A. Abdulkadir, Y. Marrakchi, A. Böhm, J. Deubner, Z. Jäckel, K. Seiwal et al., "U-net: deep learning for cell counting, detection, and morphometry," *Nature methods*, vol. 16, no. 1, pp.67–70, 2019.

[30]    A. H. Abdi, S. Kasaei, and M. Mehdizadeh, "Automatic segmentation of mandible in panoramic x-ray," *Journal of Medical Imaging*, vol. 2, no. 4, p. 044003, 2015.

[31]    K. He, X. Zhang, S. Ren, and J. Sun, "Delving deep into rectifiers: Surpassing human-level performance on imagene classification," *In Proceedings of the IEEE international conference on computer vision*, 2015, pp. 1026–1034.

[32]    D. P. Kingma and J. Ba, "Adam: A method for stochastic optimization," in *Proceedings of the 3rd international conference for learning representations (ICLR'15)*, 2015.

[33]    N. Otsu, "A threshold selection method from gray-level histograms," *IEEE transactions on systems, man, and cybernetics*, vol. 9, no. 1, pp. 62–66, 1979.

[34]    F. Wilcoxon, "Individual comparisons by ranking methods," *Biometrics Bulletin*, vol. 1, no. 6, pp. 80–83, 1945.

[35]    M. Sepehrian, A. M. Deylami, and R. A. Zoroofi, "Individual teeth segmentation in cbct and msct dental images using watershed," in *2013 20th Iranian Conference on Biomedical Engineering (ICBME)*, 2013, pp. 27–30.